\documentstyle[aps,multicol,epsf,epsfig]{revtex}

\begin{document}

\draft
%\tightenlines
 
\title{Fluctuations and correlations in sandpile models}

\author{Alain Barrat$^{1}$, Alessandro Vespignani$^{2}$, and 
Stefano Zapperi$^{3}$}

\address{
$^1$ Laboratoire de Physique Th\'eorique, B\^atiment 210, Universit\'e 
de Paris-Sud, 91405 Orsay Cedex, France\cite{umr}\\
$^2$ The Abdus Salam International Centre for Theoretical Physics (ICTP) 
P.O. Box 586, 34100 Trieste, Italy\\
$^3$PMMH-ESPCI,
10, rue Vauquelin, 75231 Paris Cedex 05, France}
 
\date{\today}

\maketitle
\begin{abstract}
We perform numerical simulations of the sandpile model
for non-vanishing driving fields $h$ and dissipation rates $\epsilon$. 
Unlike simulations performed in the slow driving limit, 
the unique time scale present in our system
allows us to measure unambiguously response 
and correlation functions. We discuss the
dynamic scaling of the model and  show that fluctuation-dissipation 
relations are not obeyed in this system.
\end{abstract}

\pacs{PACS numbers: 05.65.+b, 05.70.Ln}

\begin{multicols}{2}

The sandpile automaton \cite{btw} is one of the simplest 
model of avalanche transport, a phenomenon of 
growing experimental and theoretical interest. 
In the model introduced by Bak, Tang and Wiesenfeld (BTW)
\cite{btw}, grains of ``energy'' are injected into the system.
Open boundary conditions \cite{btw} or bulk dissipation 
insure a balance between input and output flow and allow for
a non-equilibrium stationary state. 
In the limit of slow external driving 
and small dissipation, which corresponds to
an infinite time scale separation between 
driving and response,
the model displays an highly fluctuating avalanche behavior,
indicative of a critical point.
Despite the impressive theoretical effort devoted to
understanding the critical behavior of the model \cite{dhar,priezz,ld},
several important issues still remain to be addressed. 

Numerical simulations are usually performed 
under slow driving and boundary dissipation, since
the limit of infinite time scale separation is easily implemented
in the computer and provides a simple way to access the avalanche critical
behavior \cite{grasma,manna,lubeck,csvz,cmvz}.
However, due to the presence of two infinitely 
separated time scales, an unambiguous definition of dynamic 
response and correlation functions is not possible \cite{hk}. 
This hinders a clear characterization of the non-equilibrium
stationary state in terms of static and dynamic response and
correlation functions.
Evaluation of these quantities helps to
elucidate the nature of the critical point and provides a test
of fluctuation-dissipation relations, at least
in some weaker sense.
Recently, it has been proposed to interpret the behavior
of sandpile models in analogy with other non-equilibrium critical 
phenomena, such as absorbing phase transitions \cite{vdmz}, 
driven interfaces in random media \cite{mid,pb,lau} 
and branching processes \cite{zl}. 
These theoretical studies suggest new ways to perform 
numerical simulations in which a unique time
scale is considered \cite{vdmz,lau,vz}.

In this letter, 
we present numerical simulations of the sandpile model for different
driving rates $h$ and study how the system approaches the 
critical point when $h\to 0$. 
In this way, we are able to measure quantities that
are not accessible in the time scale separation 
regime. The local density of active sites, that can be identified as the
order parameter of the model \cite{vz}, is homogeneous only in the case of bulk
dissipation. For boundary dissipation, it displays a
marked curvature, that was
anticipated in Refs.\cite{vdmz,lau}
and could explain several scaling anomalies found in the BTW model.
The energy landscape is instead homogeneous in both cases and
its statistical properties do not depend on the 
dissipation rate $\epsilon$ in the limit $h\to 0$.

We measure correlation and response functions in time and space domains
and observe the scaling of the related characteristic lengths and times. 
We find two different characteristic times, 
implying that fluctuation-dissipation relations are not obeyed.
We observe, however, a well defined scaling behavior 
and the value of the critical exponents
are in agreement with recent large scale numerical 
simulations of slowly driven sandpiles \cite{lubeck,csvz,cmvz}.
Finally, the present numerical analysis opens the road to future
studies to resolve some longstanding problems such as
the precise identification of universality classes for these models
\cite{csvz}.  

In sandpile models \cite{btw},
each site $i$ of a $d-$dimensional lattice bears an integer variable 
$z_i \geq 0$, which we call {\em energy}. 
At each time step an energy grain is added on a randomly chosen site.
When a site reaches or exceeds a threshold
$z_c$ it topples: $z_i \to z_i-z_c$, and $z_j \to z_j+1$ 
at each of the $g$ nearest neighbors (nn) of $i$. Each toppling can trigger
nn to topple and so on, generating an avalanche. The original
BTW model is conservative and energy is dissipated only at the boundary, i.e.
energy grains from toppling boundary sites flow out of the system.
Infinitely slow driving is implicitly built into the model: during 
the avalanche the energy input stops, until the system is again 
quiescent (no active sites are present), so that we can identify
two distinct time scales $T_d$ and $T_a$, for driving 
and activity, respectively.  A single driving time 
step can in principle be followed by an
infinite number of avalanche time steps and
$T_a/T_d \to 0$. For this reason, there are
two possible definitions for the correlation function, depending on the 
choice of the scale used to measure time (slow or fast) \cite{hk,giaco}.

Here we simulate the BTW sandpile model for  
a non-vanishing  driving field: each site has a probability $h$
per unit time to receive an energy grain, also if active sites 
are present in the system. This defines a unique time step for both 
driving and activity updating. The parameter $h$ sets the 
driving rate, and in the limit $h\to 0^+$ we recover the slow 
driving limit; i.e. during the evolution of an avalanche the system
does not receive energy. We consider two possible mechanisms for 
energy dissipation: {\it (i)} usual boundary dissipation
and {\it (ii)} bulk dissipation, simulated  
introducing the probability $\alpha$ that a
toppling site looses its energy without transferring it to the
neighbors,
which corresponds to an effective average dissipation
$\epsilon=\alpha z_c$. In case (ii), we impose periodic boundary
conditions. We use two dimensional lattices with linear sizes
ranging from $L=64$ to $L=300$, and parameters in the range
$10^{-6}<h<10^{-3}$ and $10^{-3}<\epsilon<10^{-2}$.

The order parameter in sandpile models is the density $\langle\rho_a\rangle$ 
\cite{braket} of active sites, whose energy is larger 
than $z_c$ \cite{btw,vdmz,mid,lau,vz}. 
The dependence of the order parameter on the control parameters 
$h$ and $\epsilon$ is readily obtained by means of 
conservation arguments \cite{vz}: 
since energy is conserved in the stationary state, the incoming energy 
flux $\langle J_{in}\rangle =hL^d$ must be equal to the dissipated energy 
$\langle J_{out}\rangle =\epsilon\langle \rho_a\rangle  L^d$. 
By equating the  two fluxes we obtain
 $\langle \rho_a\rangle =h/\epsilon$. In systems with boundary 
dissipation, the effective dissipation rate scales with the system size as 
$\epsilon\sim L^{-\mu}$, with $\mu=2$ \cite{vz}, 
yielding $\langle \rho_a\rangle \sim h L^2$. It has to be noticed 
that the model is critical just in the double limit $\epsilon\to 0$ and 
$h/\epsilon\to 0$. The onset of a nonvanishing driving thus destroys 
criticality in that it enforces a nonzero dissipation. For $h\ll\epsilon$
the cutoff length scaling is dominated only by dissipation, while 
for greater driving fields more complicate scaling behaviors occur\cite{vz}. 

We study the behavior of the density of active sites 
in the system and measure the stationary 
average density of local energies; i.e. the density $\rho_i$ of sites 
with $i$ energy grains. In Fig.~1 we report the behavior
of the densities as a function of $h/\epsilon$. 
For small values of $h/\epsilon$ we find  
$\langle \rho_i\rangle =\rho_i^{0}+c_ih/\epsilon +{\cal
  O}((h/\epsilon)^2)$, where $\rho_i^{0}$ are 
the values extrapolated from the limit $h\to 0^+$ and are
given by  $\rho_0^0=0.075(1)$, $\rho_1^0=0.176(1)$, 
$\rho_2^0=0.307(1)$ and $\rho_3^0=0.442(1)$. These values are 
in excellent agreement with the exact  values obtained for the slowly 
driven sandpile (with boundary dissipation)
\cite{priezz} and are independent of the dissipation rate. 
This implies that the energy substrate over which avalanches 
propagate is the same in the case of bulk and boundary dissipation.  
For $i > 3$ we  obtain $\rho_i^{0}=0$ and for small 
$h$ we observe $\langle \rho_a\rangle \simeq \langle \rho_4\rangle $, 
while for larger $h$ higher energy 
levels become populated and $\langle \rho_a\rangle $ has non negligible 
contributions coming from $\langle \rho_i\rangle $ 
with $i>4$. Finally we confirm that $\langle \rho_a\rangle = h/\epsilon$
for the whole range of parameters.
In the case of boundary dissipation we 
recover  $\langle \rho_a\rangle \sim h L^2$.

To elucidate the differences between bulk and boundary 
dissipation, we measure the local density of active sites $\langle
\rho_a(r)\rangle$. In the case of bulk dissipation the 
density profile is flat
$\langle \rho_a(r)\rangle =\langle \rho_a\rangle $, while 
in the case of boundary dissipation we 
obtain a surface that can be well approximated by a paraboloid 
(see Fig.~2). This is due to the 
highly inhomogeneous dissipation  which imposes a zero density 
of active sites on the lattice boundary and corresponds to an 
elastic interface pinned 
at the boundaries as discussed in Ref.s~\cite{vdmz,lau}.
This effect can explain the anomalies encountered in the numerical 
evaluation of avalanche exponents \cite{lubeck,cmvz} and
the persistent deviations from simple scaling 
observed in BTW model \cite{att}. 

In order to obtain a quantitative description of the stationary state, 
we study the effect  on the 
stationary density of a small perturbation in the driving field
\begin{equation}
\Delta\rho_a(r,t)=\int\chi_{h,\epsilon}(r-r',t-t')\Delta h(r',t')dr' dt'
\end{equation} 
where $\chi_{h,\epsilon}(r,t)$ is the local response function.
In the limit $h\to 0^+$ the integrated 
susceptibility $\chi\equiv\int dtd^dr \chi_{h,\epsilon}(r,t)$
scales as the average avalanche size $\chi\sim \langle s\rangle $ 
and the time integrated response function scales as \cite{vz}
\begin{equation}
\bar{\chi}_{h\to 0,\epsilon}(r) \equiv
\int\chi_{h\to 0,\epsilon}(r,t) dt \propto \frac{1}{r^{d-2}}e^{-r/\xi}
\label{chi_int}
\end{equation} 
where $\xi$ is the characteristic length. Since
$\chi=\partial\rho_a/\partial h$ and $\rho_a=h/\epsilon$,
the response function diverges in the limit of 
vanishing driving and dissipation as $\chi = 1/\epsilon$. 
By noting that $\chi\sim \xi^2$, we obtain that 
$\xi\sim\epsilon^{-\nu}$ with $\nu=1/2$ \cite{vz}. 
These results can be obtained in mean-field (MF) theory 
but hold in all dimensions due to conservation \cite{vdmz,lau,vz}.

To measure the response function, we drive the system
in the stationary state with a given $h$ and we then add $n$ energy
grains (i.e. $\Delta h=n/L^2$)
on a given lattice site\cite{notapert}.
The time integrated response function is equivalent 
to the average difference of activity 
$\bar{\chi}_{h,\epsilon}(r)=
\Delta\rho_a(r)=\langle \rho_a(r)\rangle_{h+\Delta h}
-\langle \rho_a(r)\rangle_{h}$, where $r$ denotes the distance from 
the perturbed site.
We observe that this function decays exponentially as predicted by 
Eq.~\ref{chi_int}
and measure the correlation length $\xi$ (see Fig.~3). 
In the case of bulk dissipation,
for small driving fields the $\xi$ depends only on 
the dissipation rate and scales as
$\xi\sim\epsilon^{-\nu}$, with $\nu=0.50\pm 0.01$ (see Fig.~4).
In the case of boundary dissipation, to evaluate 
$\bar{\chi}_{h,L}(r)$ we have to consider explicitly the spatial inhomogeneity
of the stationary density: $\langle \rho_a(r)\rangle \neq h/\epsilon $.
We observe also in this case that the integrated response function 
decays exponentially and defines a correlation length increasing
linearly with the lattice size; i.e $\xi\sim L$. 
This result does not agree with the 
anomalous scaling found in a continuous energy sandpile\cite{giaco}.
We perform analogous simulations in $d=3$ and find that Eq.~\ref{chi_int}
is still verified \cite{bvz}. 

Furthermore, we study the response function in the time domain  defined as 
$\tilde{\chi}_{h,\epsilon}(t)\equiv\int dr \chi_{h,\epsilon}(r,t)$ after a 
small variation $\Delta h$ of the driving field.
Also in this case we obtain a clear exponential
behavior defining the characteristic time scale $\tau$.
For small driving field, $\tau$ scales as a function
of the dissipation rate as $\tau\sim\epsilon ^{-\Delta}$, with 
$\Delta=0.75\pm0.05$ (see Fig.~4). We then 
evaluate the dynamical exponent $z=\Delta/\nu=1.5\pm0.1$ relating 
time and spatial characteristic length: $\tau \sim \xi^z$. 
In the limit $h\to 0^+$, we expect that the critical exponents $\nu$
and $z$ express the divergence of avalanche characteristic size and
time, respectively. 
The numerical results confirm this observation \cite{cmvz}.
For increasing values of $h$, the driving field enters the scaling
form and the results will be reported elsewhere\cite{bvz}.

We now turn to the analysis of the correlation function defined as
$C(r,t)=\langle \rho_a(r,t)\rho_a(0,0)\rangle -\langle \rho_a\rangle^2$.
In previous simulations, performed in the slow driving limit, 
correlation functions were usually measured with respect to the 
slow time scale \cite{btw,giaco,brigita} and the fast time scale was explored 
studying the avalanche propagation.
The introduction of non vanishing driving and dissipation allows us 
to bridge the gap between  the two regimes.
We study the correlation function 
in time and space domains and find an exponential 
decay at long times and distances \cite{note3}, defining the correlation 
lengths $\xi_c$ and $\tau_c$ for space and time, respectively. 
The scaling of these correlation lengths is in agreement
with the one obtained analyzing the response functions
(i.e. $\xi_c \sim \epsilon^{-\nu}$, with $\nu\simeq0.5 $ and 
$\tau_c \sim \epsilon^{-\Delta}$, with $\Delta\simeq 0.75$) and 
confirms the existence of a unique critical behavior in time and space
(see Fig.~4).

In order to clarify the interplay  
between slow and fast dynamical modes, we analyze
fluctuation-dissipation relations. 
In equilibrium phenomena, the fluctuation-dissipation
theorem ensures that the response of the system to a small perturbation is
related to the correlation function. In particular, the response
function is given by
\begin{equation}
\chi(t)= -\frac{1}{T}\frac{dC (t)}{dt},
\label{eq:fdt}
\end{equation}
where $T$ is the temperature. Eq.~\ref{eq:fdt} is strictly verified
only in equilibrium systems, but it has been recently generalized
to some classes of non equilibrium systems, namely
systems displaying ``aging'' \cite{fdt}. In those examples the 
fluctuation-dissipation relation provides an information on 
an effective non equilibrium temperature that rules the dynamical
evolution of the system. 

We test Eq.~\ref{eq:fdt} and we find that the usual linear behavior 
does not hold. On the contrary, we show that the parametric plot 
of $\chi(t)$ versus $C(t)$  
defines a   power law behavior, as
shown in the double logarithmic
plot of  Fig.~5. This is striking evidence that the
fluctuation-dissipation relation does not hold in these systems.
Since we are in presence of two exponential functions, 
the linear behavior on the logarithmic scale 
is the signature of two different  values for 
characteristic times $\tau_c$
and $\tau$ for the correlation and response function respectively. 
The slope indicates the ratio among the two time scales is
given by $\tau_c/\tau\simeq 0.4$ and does not depend on
driving and dissipation rates. This observation reflects
the fact that the correlation and the response times scale
with the same exponents with respect to dissipation and 
define unambiguously the critical behavior of the model. 
In particular, it implies that
the dynamical exponent $z\simeq1.5$ is unique and can be estimated 
either by measuring avalanche distributions or the
correlation functions. 
Previous simulations \cite{lubeck,giaco} revealed two
different dynamical exponents in the fast and in the slow time scale.
These differences are probably due to 
the ambiguous  definition of time in the infinite 
time scale separation limit.

Finally, we note that it is not possible to define an effective
temperature for the dynamics of sandpile models.
It is interesting to compare this observation with a recent
work \cite{rundle} showing that
the stationary state of non-equilibrium threshold models, similar
to the one studied here, can be described by Boltzmann statistics 
in the mean-field limit. The validity of claim of Ref.~\cite{rundle}
has been debated in the literature \cite{sor-run}. 
We measure fluctuation-dissipation relations in a
random neighbor sandpile model, which is described by mean-field
theory, and find that fluctuation-dissipation 
relations are not satisfied \cite{bvz},
in  disagreement with the conclusions of Ref.~\cite{rundle}.

We thank A.Chessa, D.Dhar, R. Dickman, S. Franz, K.B. Lauritsen,   
E.Marinari, M. A. Mu\~noz, R. Pastor-Satorras and A. Stella 
for comments and discussions. A.V. and S.Z. acknowledge partial support 
from the European Network Contract ERBFMRXCT980183.

%\newpage
\narrowtext
\begin{figure}
\centerline{
      \epsfig{figure=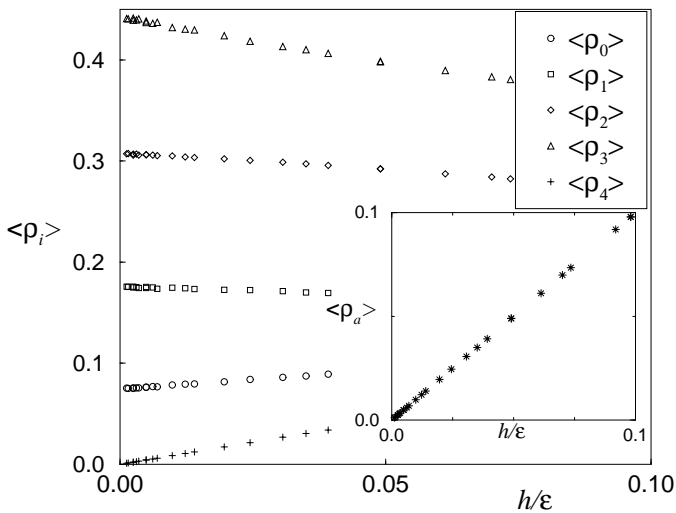,width=7cm,angle=-90}
       }
\caption{Mean densities $\langle \rho_i \rangle$ of sites with 
	 $i$ energy grains, vs $h/\epsilon$; Inset: mean density 
	 $\langle \rho_a \rangle$ of active sites,
         vs $h/\epsilon$. Values of $h$ range from $10^{-6}$ to
         $10^{-3}$, and $\epsilon$ is between $10^{-3}$ and 
	 $10^{-2}$. We observe that $\langle \rho_a \rangle=h/\epsilon$,
         and that the various $\langle \rho_i \rangle$
	 depend on $h$ and $\epsilon$ only through the ratio 
	 $h/\epsilon$ (see text).} 
\label{fig:1}
\end{figure}

\begin{figure}[bt]
\centerline{
       \epsfig{figure=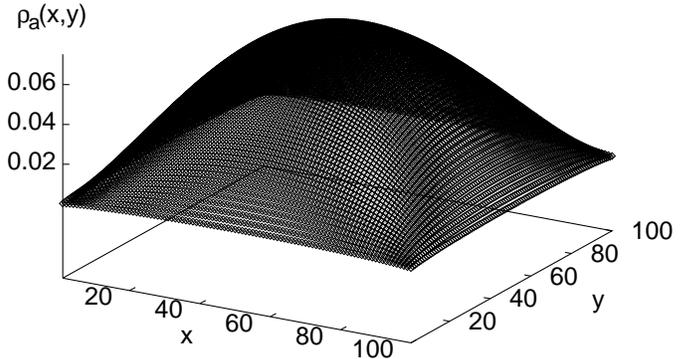,width=7cm,angle=-90}
       }
\caption{Local density of active sites $\langle \rho_a(x,y)\rangle$,
         in the case of boundary dissipation; the linear lattice size 
         is $L= 100$, and $h=10^{-4}$.}
\label{fig:2}
\end{figure}

\begin{figure}[bt]
\centerline{
        \epsfig{figure=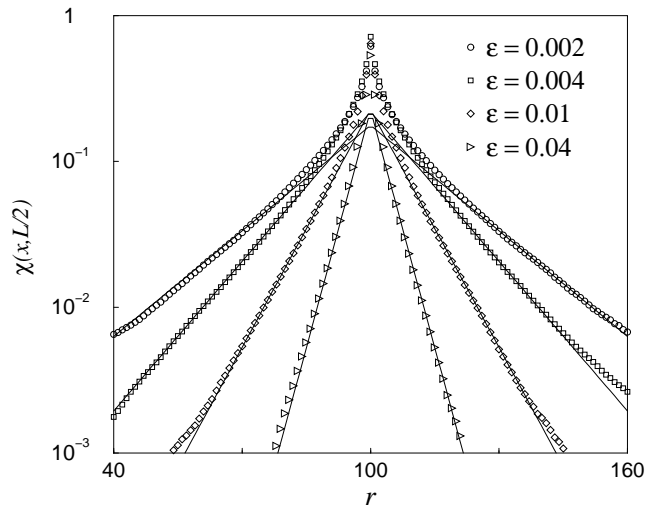,width=7cm,angle=-90}
        }
\caption{Time integrated response function  
         $\bar{\chi}_{h\to 0,\epsilon}$
         to a constant perturbation as a function of  $r$; 
         the linear lattice size is $L= 200$. The lines are exponential fits.}
\label{fig:3}
\end{figure}

\begin{figure}[bt]
\centerline{
        \epsfig{figure=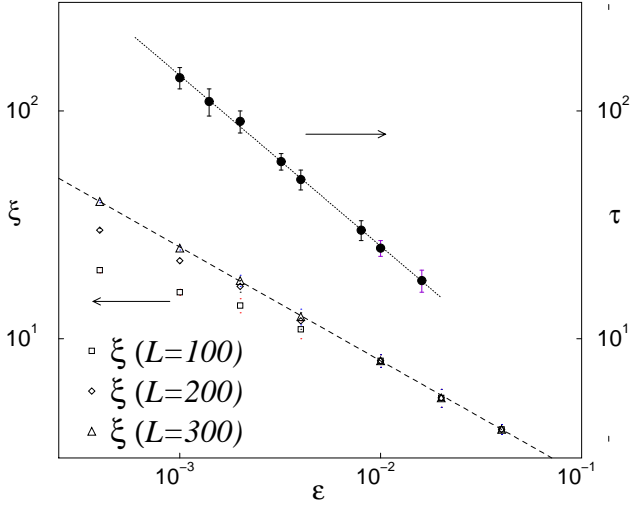,width=7cm,angle=-90}
        }
\caption{Characteristic length $\xi$ and characteristic time
         $\lim_{h \to 0^+} \tau(h,\epsilon)$ estimated from 
         the spatial response and correlation functions with bulk
         dissipation and for various system sizes and dissipation rates. 
	 For small dissipation 
	 rates $\xi\to \infty$, and larger lattice sizes must be used. 
	 The straight lines represent the best fits with slope $\nu=0.5$ and 
	 $\Delta=0.75$ for $\xi$ and $\tau$ respectively.}
\label{fig:4}
\end{figure}

\begin{figure}[bt]
\centerline{
        \epsfig{figure=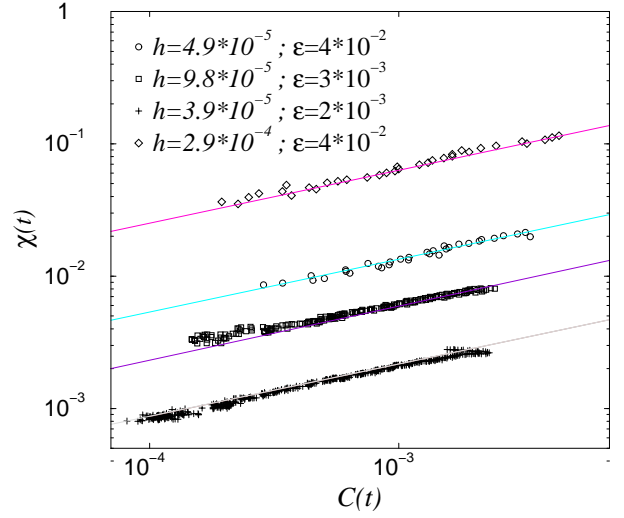,width=7cm,angle=-90}
        }
\caption{Double logarithmic plot of $\chi(t)$ vs $C(t)$ for various
         values of $h$ and $\epsilon$; the slope of the
         straight lines is  $\tau_c/\tau \simeq 0.4$.}
\label{fig:5}
\end{figure}

\end{multicols}


\begin{thebibliography}{99}
\bibitem{note3}
        The correlation function sometimes shows  periodic
        oscillations that can 
        be described in mean-field theory \cite{vz}, and will be 
        reported in \cite{bvz}.
\end{thebibliography}
\end{document}